\documentclass[a4paper,10pt,twoside]{cpc-hepnp}
\usepackage{CJK,upgreek,fancyhdr}
\usepackage{multicol}
\usepackage{graphicx}
\usepackage{color}
\usepackage{booktabs}
\usepackage{amssymb,bm,mathrsfs,bbm,amscd}
\usepackage[tbtags]{amsmath}
\usepackage{lastpage}
\newcommand{\be}{\begin{equation}}
\newcommand{\ee}{\end{equation}}
\newcommand{\edm}{\end{displaymath}}

\newcommand{\barr}{\begin{array}}
\newcommand{\earr}{\end{array}}

\DeclareMathAlphabet{\mathitbf}{OML}{cmm}{b}{it}
\begin{document}
\begin{CJK*}{GB}{gbsn}
\fancyhead[c]{\small Chinese Physics C~~~Vol. xx, No. x (201x) xxxxxx}
\fancyfoot[C]{\small 010201-\thepage}
\footnotetext[0]{Received 6 November 2016}

\title{Why static bound-state calculations of tetraquarks
should be met with scepticism}
\author{%
      George Rupp$^{1;1)}$\email{george@tecnico.ulisboa.pt}%
\quad Eef van Beveren$^{2;2)}$\email{eef@teor.fis.uc.pt}%
}
\maketitle
\address{%
$^1$ Centro de F\'{\i}sica e Engenharia de Materiais Avan\c{c}ados,
Instituto Superior T\'{e}cnico, Universidade de Lisboa, P-1049-001 Lisboa,
Portugal \\
$^2$ Centro de F\'{\i}sica da UC, Departamento de F\'{\i}sica, 
Universidade de Coimbra, P-3004-516 Coimbra, Portugal
}
\begin{abstract}
Recent experimental signals have led to a revival of tetraquarks, the
hypothetical $q^2\bar{q}^2$ hadronic states proposed by Jaffe in
1976 to explain the light scalar mesons. Mesonic structures with exotic
quantum numbers have indeed been observed recently, though a controversy
persists whether these are true resonances and not merely kinematical 
threshold enhancements, or otherwise states not of a true $q^2\bar{q}^2$ 
nature. Moreover, puzzling non-exotic mesons are also often claimed to have
a tetraquark configuration. However, the corresponding model calculations
are practically always carried out in pure and static 
bound-state approaches, ignoring
completely the coupling to asymptotic two-meson states and unitarity, 
especially the dynamical effects thereof. In this short paper we argue that
such static predictions of real tetraquark masses are highly unreliable and
provide little evidence of the very existence of such states.
\end{abstract}
\footnotetext[0]{\hspace*{-3mm}\raisebox{0.3ex}{$\scriptstyle\copyright$}201    3
Chinese Physical Society and the Institute of High Energy Physics
of the Chinese Academy of Sciences and the Institute
of Modern Physics of the Chinese Academy of Sciences and IOP Publishing Ltd}    %
\begin{multicols}{2}
\section{Introduction}
In 1976 R.~L.~Jaffe proposed \cite{PRD15p267} an ingenious construct to
explain the low masses of the light scalar mesons, in the context of the MIT
bag model \cite{PRD9p3471}. His solution amounted to introducing
``crypto-exotic'' colourless $qq\bar{q}\bar{q}$ configurations instead of the
usual $q\bar{q}$ ones for ordinary mesons. Due to the very large and attractive
colour-magnetic spin-spin interaction for the ground-state $q^2\bar{q}^2$
systems, an enormous negative mass shift could be obtained and so arrive at
reasonable masses for the lightest scalar mesons \cite{PRD15p267}, viz.\
650~MeV for the $\epsilon$ (now called $f_0(500)$
\cite{PDG2016,PR658p1} or $\sigma$), 900~MeV for the $\kappa$
($K_0^\star(800)$ \cite{PDG2016}), and 1100~MeV for both the $S^\star$
($f_0(980)$ \cite{PDG2016}) and the $\delta$ ($a_0(980)$ \cite{PDG2016}).
Although Jaffe's proposal was received with general approbation, interest
rapidly faded owing to the poor status of the light scalars in those days and
lacking experimental indications of truly exotic (necessarily non-$q\bar{q}$)
states.

Renewed interest in the $q^2\bar{q}^2$ or ``tetraquark'' \cite{NPBPS21p254}
model resulted foremostly from the experimental discovery of a number of mesons
that did not seem to fit in the traditional static quark model (SQM), which
treats hadrons as manifestly bound states of quarks and antiquarks. Indeed, the
most widely used SQM, viz.\ the relativised quark model of mesons by Godfrey
and Isgur \cite{PRD32p189}, predicted considerably higher masses for enigmatic
mesons as e.g.\ the scalar charmed-strange $D_{s0}^\star(2317)$ \cite{PDG2016}
and the axial-vector charmonium $X(3872)$ \cite{PDG2016}. Over the following
years, a large number of puzzling mesonic enhancements were observed,
most of these in the hidden-charm sector and some with hidden bottom, a few
of them even electrically charged.
The latter ones would of course exclude simple $c\bar{c}$/$b\bar{b}$
assignments, if indeed confirmed as genuine resonances, justifying speculations
that they may be tetraquarks. For some time these states were labelled ---
quite arbitrarily --- ``$X$'', ``$Y$'', or ``$Z$'', but the PDG  now calls them
all $X$s \cite{PDG2016}. For very recent reviews, see
Ref.~\cite{PR639p1} on hidden-charm pentaquark and tetraquark states,
and Ref.~\cite{FBS57p1185} on exotic hadrons in general.

Despite these exciting observations, figuring out the true nature of all
these unusual states is an enormous challenge. First of all, the experimental
identification of several $X$ structures as bona-fide resonances is anything
but undisputed. In a recent lattice calculation \cite{PRD91p014504}, with many
two-meson interpolating fields included, no evidence was found of isovector
hidden-charm tetraquark states up to 4.2~GeV. Moreover, several authors
\cite{EPL96p11002,PRD88p036008,PLB747p410,PRD91p034009,PLB748p183} interpret
charged hidden-charm or hidden-bottom signals rather as non-resonant cusp-like
structures, resulting from kinematical triangle singularities in
intermediate-state diagrams. On the other hand, even when four-quark
states are predicted in model calculations, these are often described as 
(quasi-)bound states of two mesons \cite{EPJA47p120,PRD88p034018}, with binding
due to $t$-channel meson exchange instead of colour forces among two quarks and
two antiquarks. Finally, there are also models suggesting that the observed
charged hidden-charm and/or hidden-bottom peaks may be highly excited $D_s$
($c\bar{s}$) states \cite{PLB669p156,POSp003} or light-quark axial-vectors
\cite{PRD94p014016}.

Now, if tetraquarks nonetheless exist in nature as genuine $q^2\bar{q}^2$ bound
states or resonances, the question remains how to describe them in a realistic
way, besides resorting to the lattice. This brings us inexorably to the issue
of mass shifts from unitarisation, sometimes called ``unquenching''
\cite{ARXIV160504260}, which we shall discuss in the next section. But let us
first quote the warning of Jaffe himself \cite{PRD15p267} about describing the
light scalar mesons as stable tetraquarks:
\begin{quote}
{\it First, we are confronted with mesons whose width is a substantial
     fraction of their mass. A calculation of their masses which ignores
     decay processes (as does ours) must not be taken too literally. We
     should not expect the accuracy we demanded in our treatment of
     $Q\bar{Q}$ mesons and $Q^3$ baryons.}
\end{quote}

\section{Unquenching the quark model}
A fundamental difference between strong interactions and e.g.\ electromagnetism
is that in the former case mass splittings and decay widths can be of similar
magnitude. Picking just one typical example from the PDG
\cite{PDG2016} Meson Summary Table, we see that the mass difference between
the tensor meson $f_2^\prime(1525)$ and its first radial excitation $f_2(1950)$
is about 420 MeV, while the full width of the latter resonance is
$(472\pm18)$~MeV. This has tremendous implications for spectroscopy, as was
recognised almost four decades ago by the Cornell \cite{PRD17p3090}, Helsinki
\cite{AOP123p1}, and Nijmegen \cite{PRD21p772} hadronic-physics groups. Namely,
most mesons/baryons are not merely bound $q\bar{q}$/$qqq$ states, but rather
resonances in meson-meson or meson-baryon scattering, respectively. Now,
arguments based on $S$-matrix analyticity imply that imaginary parts of 
resonance poles are in principle of the same order as the corresponding real
shifts with respect to the corresponding bound states from quark confinement
only. This may give rise to huge distortions of hadron spectra as predicted 
by the SQM. To make life worse, relatively stable hadrons, with widths of
roughly 1 MeV or even less, can still be subject to real mass shifts at
least two orders of magnitude larger, due to virtual decay. A famous example
is the enigmatic scalar charmed-strange meson $D_{s0}^\star(2317)$
\cite{PDG2016}, predicted 170--180 MeV heavier by the SQM, but ending up at
a much lower mass owing to the closed yet strongly coupling $S$-wave $DK$
decay channel \cite{PRL91p012003}. The latter model result was recently
confirmed on the lattice \cite{PRL111p222001}, thus enfeebling claims
\cite{PRD68p011501} of a tetraquark interpretation of this meson.

In order to illustrate the possible effects of unquenching on meson spectra in
general, we collect in Table~\ref{massshifts} several model calculations of
mass shifts owing to strong 
\begin{center}
\tabcaption{ \label{massshifts}
Negative real mass shifts from unquenchimg.  Abbreviations:
$P,V,S$ = pseudoscalar, vector, scalar mesons, respectively; $q$ = light quark.
See text and Ref.~\cite{ARXIV160504260} for further details.}
\footnotesize
\begin{tabular*}{80mm}{c@{\extracolsep{\fill}}cc}
\toprule
Refs.\ &
Mesons & $-\Delta M$ (MeV) \\ 
\hline
\cite{PRD17p3090} &
charmonium & 48--180  \\
\cite{AOP123p1,ZPC5p205} &
light $P$, $V$ & 530--780, 320--500 \\
\cite{PRD21p772,PRD27p1527} &
$q\bar{q}$, $c\bar{q}$, $c\bar{s}$, $c\bar{c}$, $b\bar{b}$; $P,V$ &
$\approx\!$ 30--350 \\
\cite{ZPC30p615} & 
$\sigma$, $\kappa$, $f_0(980)$, $a_0(980)$ & 510--830 \\
\cite{ZPC30p615} & 
standard $S$ (1.3--1.5 GeV) & $\sim0$ \\
\cite{PRD42p1635} & 
$\rho(770)$, $\phi(1020)$ & 328, 94 \\
\cite{PRL91p012003}  & 
$D_{s0}^\star(2317)$, $D_0^\star(2400)$ & 260, 410 \\
\cite{PRD70p114013} & 
$D_{s0}^\star(2317)$, $D_s^\star(2632)$ & 173, 51 \\
\cite{PRD72p034010} & 
charmonium & 165--228 \\
\cite{PRC77p055206} & 
charmonium & 416--521 \\
\cite{EPJC71p1762} & 
$X(3872)$ & $\approx\!100$ \\
\cite{PRD84p094020}  & 
$c\bar{q}$, $c\bar{s}$; $J^P=1^+$ & 4--13, 5--93  \\ 
\bottomrule
\end{tabular*}
\vspace{0mm}
\end{center}
\vspace{3mm}
decay. Note that not all of these approaches amount
to full-fledged $S$-matrix unitarisations of the SQM, in fact only those in
Refs.~\cite{PRD17p3090,PRD21p772,PRD27p1527,ZPC30p615,EPJC71p1762,PRD84p094020}
(for further details, see Ref.~\cite{ARXIV160504260}). But even among the
latter there can be sizable differences, as we can see in Table~\ref{massshifts}
by comparing the predictions of Refs.~\cite{PRD17p3090} and \cite{PRD21p772} for
charmonium. These disagreements not only originate in different confinement
forces, but also in the employed decay mechanisms, which are in
their turn influenced by the nodal structure of the $q\bar{q}$ wave functions.
Nevertheless, Table~\ref{massshifts} shows potentially huge mass shifts, some
of which are even larger than typical radial spacings in meson spectra. Also
note that $S$-matrix calculations generally produce complex shifts, whenever 
at least one decay channel is open. Particularly interesting in this respect is
the case of the charmed-light axial-vector meson $D_{1}(2430)$, whose
imaginary mass shift in Ref.~\cite{PRD84p094020} came out an order of
magnitude larger than its real shift, with the corresponding resonance pole 
position being in good agreement with experiment \cite{PDG2016}. Note that this
is a highly non-perturbative effect and not a consequence of the usual
perturbative calculation of the width.

The most surprising result in Table~\ref{massshifts} is, though, for the light
scalar mesons, which emerged as a complete nonet of dynamical resonances in the
30-years-old model calculation of Ref.~\cite{ZPC30p615}, without any parameter
fitting. The scalar-meson mass shifts of 510--830~MeV from unitarisation
reported in Table~\ref{massshifts} correspond to the differences between the
bare (``quenched'') $1\,{}^{3\!}P_0$ quark-antiquark energy levels and the real
parts of the lowest scalar resonance poles. However, the latter appear as extra,
dynamically generated poles, besides an also complete nonet of scalar
resonances that shift much less (cf.\ $D_{1}(2430)$ above) and remain in the
range 1.3--1.5 GeV
\cite{ZPC30p615}. This allows to describe both the light scalar nonet
$f_0(500)$ ($\sigma$), $K_0^\star(800)$ ($\kappa$), $f_0(980)$, $a_0(980)$
\cite{PDG2016} and the standard ground-state scalar nonet $f_0(1370)$,
$K_0^\star(1430)$, $f_0(1500)$, $a_0(1450)$ \cite{PDG2016} as unitarised
$q\bar{q}$ states. Recent work \cite{NPB909p418} supports this phenomenon of
generating extra resonances in the light scalar sector (also see
Ref.~\cite{ARXIV160806569} and references therein).

In the next section we shall focus on the light scalar mesons in other
approaches, keeping in mind the importance of unitarisation.

\section{Light scalar meson nonet}

As already said above, the observation of many mysterious mesonic signals over
the past decade has led to a revival of Jaffe's \cite{PRD15p267} tetraquark
model, also for the light scalars (see e.g.\ Ref.~\cite{PLB662p424}). The
problem is that practically all these works simply ignore unitarisation. This
is all the more serious in the scalar-meson case, as any postulated 
tetraquark wave function will inevitably contain components of two colourless
$q\bar{q}$ subsystems in a relative $S$-wave. So if phase space allows, such
a hypothetical tetraquark can simply fall apart into two mesons, like e.g.\
$f_0(500) \to \pi\pi$ or $K_0^\star(800) \to K\pi$, which was recognised by
Jaffe already 40 years ago (see Ref.~\cite{PRD15p267} for the mentioned
figures):
\begin{quote}
{\it If it is heavy enough an S-wave $Q^2\overline{Q}^2$ meson will be
     unstable against decay into two S-wave $Q\overline{Q}$ wave mesons.
     The $Q^2\overline{Q}^2$ state simply falls apart, or dissociates,
     as illustrated in Fig. 4(a). In contrast, decay of a $Q\overline{Q}$
     meson into two $Q\overline{Q}$ mesons (for example $\rho \to \pi\pi$
     or $f \to \pi\pi$) requires creation of a $Q\overline{Q}$ pair
     [Fig. 4(b)].}
\end{quote}
In face of this physical reality, Jaffe and Low developed \cite{PRD19p2105} 
the so-called $P$-matrix formalism, which should relate $S$-matrix observables,
with boundary conditions at infinity, to solutions of a relativistic wave
equation with boundary conditions at an arbitrary finite distance and the
corresponding discrete energy levels. They then applied it to $S$-wave
meson-meson scattering, extracting real energies corresponding to $P$-matrix
poles from experimental scattering data and comparing these to the
MIT-bag-model mass predictions \cite{PRD15p267} for light scalar tetraquarks.
However, it remains unclear how to justify a direct link, in a single-channel
approach, between asymptotic two-meson states and a wave function for 4
coloured quarks confined to a bag. Also, no quark-antiquark annihilation is
considered in this formalism, which the 
authors themselves expected \cite{PRD19p2105} to occur in mixing of heavier
$q^2\bar{q}^2$ and $q\bar{q}$ states. Moreover, the $P$-matrix method does
not allow to draw conclusions on resonance widths from the data. Finally,
the experimental data do not support the existence of exotic or
crypto-exotic states between 1 and 2 GeV that should correspond to the
$P$-matrix poles of heavier scalar tetraquark bag states predicted in
Ref.~\cite{PRD15p267}.

In principle, the dynamical consequences
of uninhibited decay may be dramatic. Suffice it to recall the enormous
mass shifts for the light scalar mesons in the unitarised $q\bar{q}$ model of
Ref.~\cite{ZPC30p615}, despite the necessity of creating a new $q\bar{q}$ pair.
An important hint may come from Ref.~\cite{PRD83p094010}, in which a unitarised
toy model of tetraquarks was formulated via a two-variable Schr\"{o}dinger
equation for the two spatial configurations $(qq)(\bar{q}\bar{q})$ and
$(q\bar{q})(q\bar{q})$. In spite of the implemented simplifications, a very
striking conclusion emerges from this study, namely that no tetraquark bound
state or observable resonance is found for zero orbital angular momentum, i.e.,
precisely in the case of scalar mesons. What may happen here is that a 
bound-state pole corresponding to a static scalar tetraquark state  moves
very far away or even disappears completely in the scattering continuum
once decay into two mesons is allowed. In order to better investigate such a
scenario, a more realistic version of the referred toy model would be
highly desirable, perhaps along the lines of Ref.~\cite{ARXIV150904943}, but
applied to a light scalar tetraquark instead of the studied $qq\bar{Q}\bar{Q}$
systems, where $Q$ stands for heavy quark ($c$ or $b$).

To conclude our discussion, we turn to a very recent \cite{PLB753p282}, 
alternative description of light scalar tetraquarks, which amounts to the
numerical solution of a four-body Bethe-Salpeter (BS) equation with pairwise
rainbow-ladder gluonic interactions. The authors formally write down a
scattering equation for the $qq\bar{q}\bar{q}$ $T$-matrix, i.e.,
\begin{equation}
T\;=\;K+K\,G_0\,T \; ,
\label{tmatrix}
\end{equation}
where $K$ is the $qq\bar{q}\bar{q}$ interaction kernel and $G_0$ the
product of four dressed (anti-)quark propagators. However, in order to search
for poles, Eq.~(\ref{tmatrix}) is immediately replaced by a homogeneous
equation for the BS amplitude (or vertex function) $\Gamma$:
\begin{equation}
\Gamma\;=\;K\,G_0\,\Gamma \; .
\label{vertex}
\end{equation}
The latter equation allows to find bound-state solutions, but in principle
it can also be used to describe a resonance for complex energy, provided that
a proper analytic continuation into the second Riemann sheet is carried out,
so as to include the corresponding pole contribution. This was done in
Ref.~\cite{PRD52p2690} for a three-dimensional relativistic two-body equation,
using contour-rotation techniques. However, doing something similar in the
four-dimensional four-body BS case must be a gargantuan enterprise, as the
authors of Ref.~\cite{PLB753p282} themselves admit:
\begin{quote} \em
This is, however, a rather formidable task which has not even been
accomplished in simpler systems so far.  \em
\end{quote}
Yet, the implications of this understandable restriction to pure bound states
could be much more far-reaching than the truncation of the kernel to pairwise
interactions only. So let us see what the lattice has to say. Quite ideal would
be to have full-fledged lattice simulations of $S$-wave $\pi\pi$,
$K\pi$, and $\eta\pi$ scattering, with meson-meson and either $q\bar{q}$ or
$qq\bar{q}\bar{q}$ interpolators included. Very recent lattice
work has tried to describe these systems with quark-antiquark and
two-meson degrees of freedom included, viz.\ the $f_0(500)$/$\sigma$
\cite{ARXIV160705900}, $K_0^\star(800)$/$\kappa$ \cite{PRD91p054008}, and
$a_0(980)$ \cite{PRD93p094506}, though with still unphysically large pion
masses. Nevertheless, the results suggest that no tetraquark configurations
are required in the description of the light scalars. On the other hand, there
are lattice indications of the importance to account for scattering
solutions, albeit for a different system.
In Ref.~\cite{POS118} several excited meson and baryon spectra
were presented, resulting from unquenched lattice calculations with fully
dynamical quarks, but without considering meson-meson scattering solutions.
Among these was a mass prediction above 1.6~GeV for the first radial
excitation of the $K^\star(892)$ resonance. But almost simultaneously, the
same lattice group published \cite{PRD88p054508} results on $P$-wave $K\pi$
scattering, employing both $q\bar{q}$ and meson-meson interpolators. This
allowed to reasonably reproduce mass and $K\pi$ decay coupling of
$K^\star(892)$, besides extracting a tentative mass for its first radial
recurrence at $(1.33\pm0.02)$~GeV, more or less compatible with the
observed \cite{PDG2016} Breit-Wigner mass of the broad $K^\star(1410)$ 
resonance. Apart from being much closer to the experimental value than the
predictions of mainstream quark models, the value of 1.33 GeV is about
300 MeV lower than the above lattice bound-state result \cite{POS118}.
So open meson-meson channels can yield a huge mass shift, even for
a radially excited meson decaying in a $P$-wave, for which unitarisation
is usually supposed to be of limited importance. One can only guess how 
large such effects might be for a ground-state tetraquark that can freely
fall apart into an $S$-wave $\pi\pi$ or $K\pi$ state.

Some final words are due concerning the results of Ref.~\cite{PLB753p282},
which may be very relevant for QCD even if having little bearing upon
experiment. The most important conclusion seems to be:
\begin{quote} \em
\ldots\ these tetraquarks are not diquark-antidiquark states but
predominantly `meson molecules' \ldots \em
\end{quote}
In other words, two colourless $q\bar{q}$ systems constitute by far the most
important component of the computed bound-state wave function, so that the
designation ``meson-meson molecule'' appears much more appropriate for such
a system than ``tetraquark''. Moreover, this microscopic modelling of the
light scalar mesons is not so different from the effective approach in
Ref.~\cite{PRD52p2690}, where only meson-meson interactions were considered,
though in a fully unitary formalism. Curiously, the latter paper reported a
very light $\sigma$-like pole in the $\pi\pi$ $S$-wave, with a real part of
387~MeV and an enormous imaginary part of 305~MeV. This former value of
387~MeV is not very far from the bound-state ``$\sigma$'' mass of 348~MeV
found in Ref.~\cite{PLB753p282}. Of course, this may be just a coincidence,
in view of the strong unitarisation effects leading to a width of more than
600~MeV in Ref.~\cite{PRD52p2690}.

\section{Conclusions}

The hypothetical tetraquark system is a hotly disputed topic in hadronic
physics nowadays. Many are led to take recent experimental signals, 
especially in the charmonium sector, as proof of their existence. However,
several alternative explanations exist, such as non-resonant structures due to
kinematical singularities, meson-meson molecules bound by $t$-channel
exchanges, and highly excited regular $q\bar{q}$ mesons. The experimental 
challenge is not only to provide high-statistics data with unambiguous
resonance characteristics, but also to find unmistakable partner states as
predicted by the tetraquark model, either in the same isomultiplet or the same
flavour multiplet.

On the theoretical side, things are by no means easier. As we hope to have
made clear, predictions of tetraquark models that ignore the effects of decay
should not be trusted. And this applies not only to states above two-meson
thresholds but also to seemingly genuine $qq\bar{q}\bar{q}$ bound states, since
the virtual meson loops corresponding to closed yet nearby thresholds will
inevitably have a significant influence, especially in the case of $S$-waves. 

As for the light scalar resonances, of which $f_0(500)$ ($\sigma$) and 
$K_0^\star(800)$ ($\kappa$) are very broad, a pure bound-state tetraquark
description cannot be realistic, although the 4-body BS calculation of
Ref.~\cite{PLB753p282} appears to approximately account for some meson-meson
contributions. Unitarising such an approach would be a huge step foreward, but
does not seem feasible for the time being. An alternative is the unitarised
tetraquark potential model of Ref.~\cite{ARXIV150904943}, but applied to the
crypto-exotic light scalars instead of exotic $qq\bar{Q}\bar{Q}$ systems.
However, also quark-antiquark annihilation would then have to be considered,
making an already very difficult problem even more cumbersome. Finally, the
lattice will probably be able at some point to settle the tetraquark issue for
the light scalar mesons, perhaps already quite soon, in view of the progress
made recently (see
Refs.~\cite{ARXIV160705900,PRD91p054008,PRD93p094506}), also on describing
vector-meson resonances (see e.g.\ Ref.~\cite{PRD88p054508}). Still, the
complication of flavour mixing and coupled $\pi\pi$-$K\bar{K}$
channels in the isosinglet $f_0(500)$ case suggests that the first scalar to
reproduce in full glory is
$K_0^\star(800)$. This would also help to finally convince the Particle Data
Group to include the latter resonance in the Meson Summary Table
\cite{PDG2016}, which we believe is overdue \cite{Barcelona2004}. We are also
convinced that such a lattice calculation will lend support to the decades-old
\cite{ZPC30p615} picture of the light scalars as dynamical resonances generated
by the strong coupling of much heavier bare $^{3\!}P_0$ $q\bar{q}$ states to
low-mass $S$-wave meson-meson decay channels. The ``tetraquark'' interpretation
of these solutions then essentially boils down to the dominant meson-meson
components apparently observed in Ref.~\cite{PLB753p282}, with a subdominant
$q\bar{q}$ component constituting the missing ingredient, besides the
restriction to real energies only. The absence of a $q\bar{q}$ component
also in Ref.~\cite{PRD52p2690} may explain the too low $\sigma$ mass and much
too large $\sigma$ width obtained in this pure meson-meson model. 
\end{multicols}
\vspace{-1mm}
\centerline{\rule{80mm}{0.1pt}}
\vspace{2mm}
\begin{multicols}{2}

\end{multicols}
\clearpage
\end{CJK*}
\end{document}